\documentclass{ws-ijmpcs}
\usepackage{graphicx}

\begin{document}

\markboth{L. Foschini}
{The unification of relativistic jets}

%
\catchline{}{}{}{}{}
%

\title{THE UNIFICATION OF RELATIVISTIC JETS}

\author{LUIGI FOSCHINI}

\address{Osservatorio Astronomico di Brera, Istituto Nazionale di Astrofisica\\ 
Via E. Bianchi 46, Merate (LC), 23807,
Italy\\
luigi.foschini@brera.inaf.it}

\maketitle

\begin{history}
\received{21 October 2013}
\end{history}

\begin{abstract}
I report about the unification of relativistic jets from compact objects. The mass range is between 1.4 and 10
billion solar masses (i.e. from neutron stars to supermassive black holes in galaxies).

\end{abstract}

\ccode{PACS numbers: 98.54.Cm, 97.80.Jp}

\section{Why jets?}	
It is today well-known that different types of cosmic sources emit jets of matter at speeds ranging from supersonic (protostars) to relativistic (Galactic binaries and Active Galactic Nuclei, AGN), and ultrarelativistic (Gamma-Ray Bursts, GRBs). Although the sources differ from each other in many details, they all share the accretion and ejection of matter organized in certain structures. Therefore, there should be a physical engine in common with all these sources (a {\it ``universal engine''}), based on simple physical laws that can work in such different environments\cite{LIVIO}. 

\section{How to unify them?}
Many theories have been proposed to explain how jets are generated. Those collecting most followers are based on the extraction of rotational energy either from the central compact object\cite{BZ} or the accretion disk\cite{BP} or both\cite{MEIER1}. Beyond the details of the individual theories and specific sources, some fundamental features are emerging. Mass, accretion rate, and spin are characterizing the central compact object, while self-similarity is important for the jet structure. Many researchers engaged themselves in the attempt to unify jets from some subclasses of sources. Just to cite a few: Mirabel and Rodr\'iguez\cite{MIRABEL} explored the connection between microquasars and GRBs; Markoff et al.\cite{MARKOFF} and Meier\cite{MEIER2} studied the scaling between microquasars and AGN; Ghisellini and Celotti\cite{GHISELLINI1} analyzed AGN, GRBs, and Galactic binaries; Nemmen et al.\cite{NEMMEN} proposed a scaling between AGN and GRBs; and many, many others. 

Perhaps, the best-known works refer to the so-called ``fundamental plane of black hole activity''\cite{MERLONI,FALCKE}. The plane pivots on the radio-X-ray correlation found both in Galactic binaries and AGN. The proposed method seems simple and powerful, because it requires just the measurement of radio and X-ray emission of a source to derive its mass. However, it is known that different physical processes can dominate the electromagnetic spectrum in the X-ray band and vary significantly depending on the type of source (either Galactic binary or AGN), thus making the correlation without physical ground. Particularly, Chiaberge\cite{CHIABERGE} has shown that also the Sun, the Moon, Jupiter, and Saturn can be well fit to the fundamental plane of black holes, although they are not spacetime singularities, obviously... Other authors (e.g. Refs.~\refcite{PARAGI,BONCHI}) have emphasized other weaknesses in such approach. In addition, in the case of Galactic binaries, a second branch emerged in the radio-X-ray correlation\cite{CORIAT,GALLO,CORBEL}, which seems to be due to high-efficiency jets from accreting neutron stars\cite{MIGLIARI}. The fundamental plane supporters continue believing in the reliability of their work and, from time to time, they revise and update their samples (e.g. Refs. \refcite{MERLONI2,KORDING,PLOTKIN}). I do not want to be overly critical with respect to the plane: their kick-off idea was good, but the selected implementation method was not appropriate. I think that it is better to tackle the problem by means of a more physical approach, based on derived measures of physical quantities, such as the jet power or the accretion disk luminosity\cite{FOSCHINI1,FOSCHINI2,FOSCHINI3}. Instead of blindly measure the electromagnetic emission in one energy band and then puzzling over the possible contributors and biases, it is better to derive immediately the interested quantities and then search for correlations as indicated by the same theory (e.g. Ref. \refcite{BK}). The advantage of a physical approach is evident when dealing with the accretion disk luminosity. Indeed, it is known that the peak frequency of the emission of a standard accretion disk is inversely proportional to the mass of the central object ($\nu_{\rm peak}\propto M^{-1/4}$): the greater is the mass, the lower is the peak frequency. Therefore, the disk luminosity can be measured at optical/UV wavelengths for AGN and at soft X-rays for Galactic binaries. I do not take into account - for the moment - GRBs and protostars, because they require a more detailed study. 

The scaling theory of a standard phenomenological jet\cite{BK} has been already successfully developed by Heinz and Sunyaev\cite{HEINZ}. Therefore, in my works, I have tried to give observational support in favor of their theory. Also in this case, other researchers have done similar works by mixing several parameters and measurements in the most different ways (e.g. Refs. \refcite{FENDER,KORDING2} in addition to the others above cited). In my previous works\cite{FOSCHINI1,FOSCHINI2,FOSCHINI3} and here, in addition to the adoption of derived physical quantities, one important novelty is given by the recent discovery of high-energy $\gamma$ rays emitted by radio-loud Narrow-Line Seyfert 1 Galaxies (RLNLS1s)\cite{LAT1,LAT2,LAT3,LAT4}. This discovery has confirmed the presence of powerful relativistic jets in such class of AGN suggested by radio observations and opened several interesting questions (for a review see Refs. \refcite{FOSCHINI4,FOSCHINI5}).

\section{What can we measure/observe?}
The sample of sources I have considered here for the scaling is composed of 53 flat-spectrum radio quasars (FSRQs)\cite{GG2009,GG2010}, 31 BL Lac Objects\cite{GG2010,TAV2010}, 15 RLNLS1s\cite{LAT3}(\footnote{See also {\tt http://tinyurl.com/gnls1s}.}), 3 stellar-mass black holes\cite{CORIAT}, and 2 neutron stars\cite{MIGLIARI}(\footnote{Data in tabular form are available at {\tt https://zenodo.org/record/7487}.}). The Galactic binaries were observed several times in different states, so that the points are 80. I started from the jet power and disk luminosity of AGN calculated by modelling the spectral energy distribution (SED)\cite{GG2009,GG2010,TAV2010}. For the sources without SED modelling, I adopted other methods. It is well-known that the radio emission of the core is a good proxy of the jet power\cite{BK}. When possible, I made use of the radio observations of the MOJAVE Project\footnote{{\tt http://www.physics.purdue.edu/astro/MOJAVE/index.html}}, which were performed by means of VLBA at 2~cm. Otherwise, I used what is available in NED\footnote{{\tt http://ned.ipac.caltech.edu/}}, with the hypothesis of $\alpha_{\rm radio}\sim 0$. For the Galactic binaries, I found what was available in the cited literature (generally, fluxes at 8.5 GHz). The conversion of radio luminosity to jet power was done by means of an empirical formula cross-calibrated with the jet power calculated by modelling the SED\cite{FOSCHINI1}. The most up-to-date formulae are:

\begin{equation}
\begin{split}
\log P_{\rm jet,radiative} & = (12\pm2) + (0.75\pm 0.04)\log L_{\rm radio,core}\\
\log P_{\rm jet,kinetic} & = (6\pm2) + (0.90\pm 0.04)\log L_{\rm radio,core} 
\end{split}
\label{eq1}
\end{equation}

The accretion disk luminosity, estimated by fitting the optical/UV emission with a multicolor blackbody model, is already available for all the blazars\cite{GG2009,GG2010,TAV2010} and a few RLNLS1s\cite{LAT3}. In the case of the missing RLNLS1s, I have first estimated the size of the broad-line region ($R_{\rm BLR}$) from the optical observations; then, I calculated the disk luminosity following the well-known $R_{\rm BLR}-L_{\rm disk}$ relationship\cite{BENTZ}. The values could be overestimated in the case of high activity of the jet, because of the synchrotron emission. However, this method has been applied to some sources only, while in all the other cases, the modelling of the SED allowed to disentangle the different contributions. 

\begin{figure}[!t]
\centering
\includegraphics[angle=270,scale=0.26]{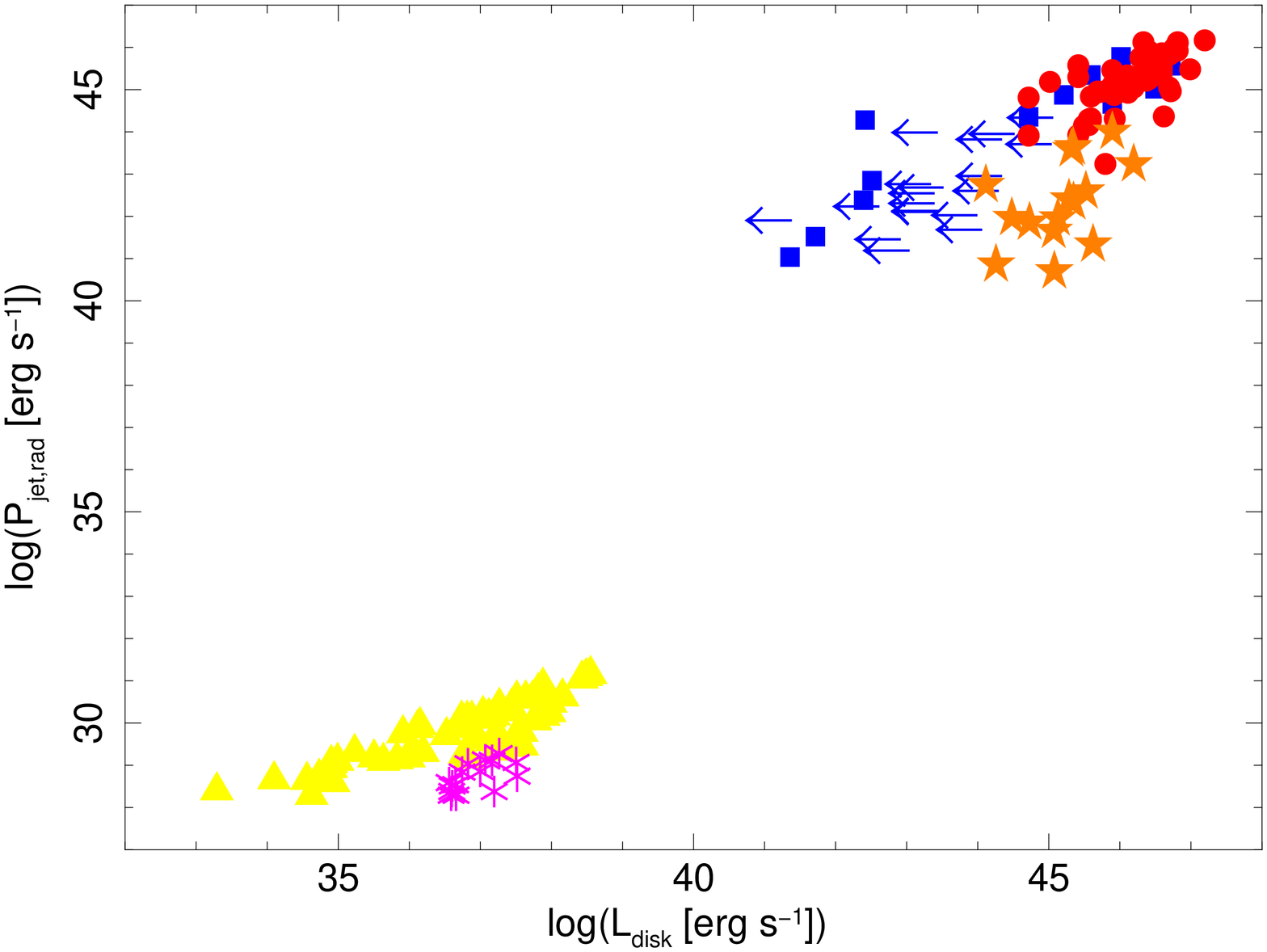}
\includegraphics[angle=270,scale=0.26]{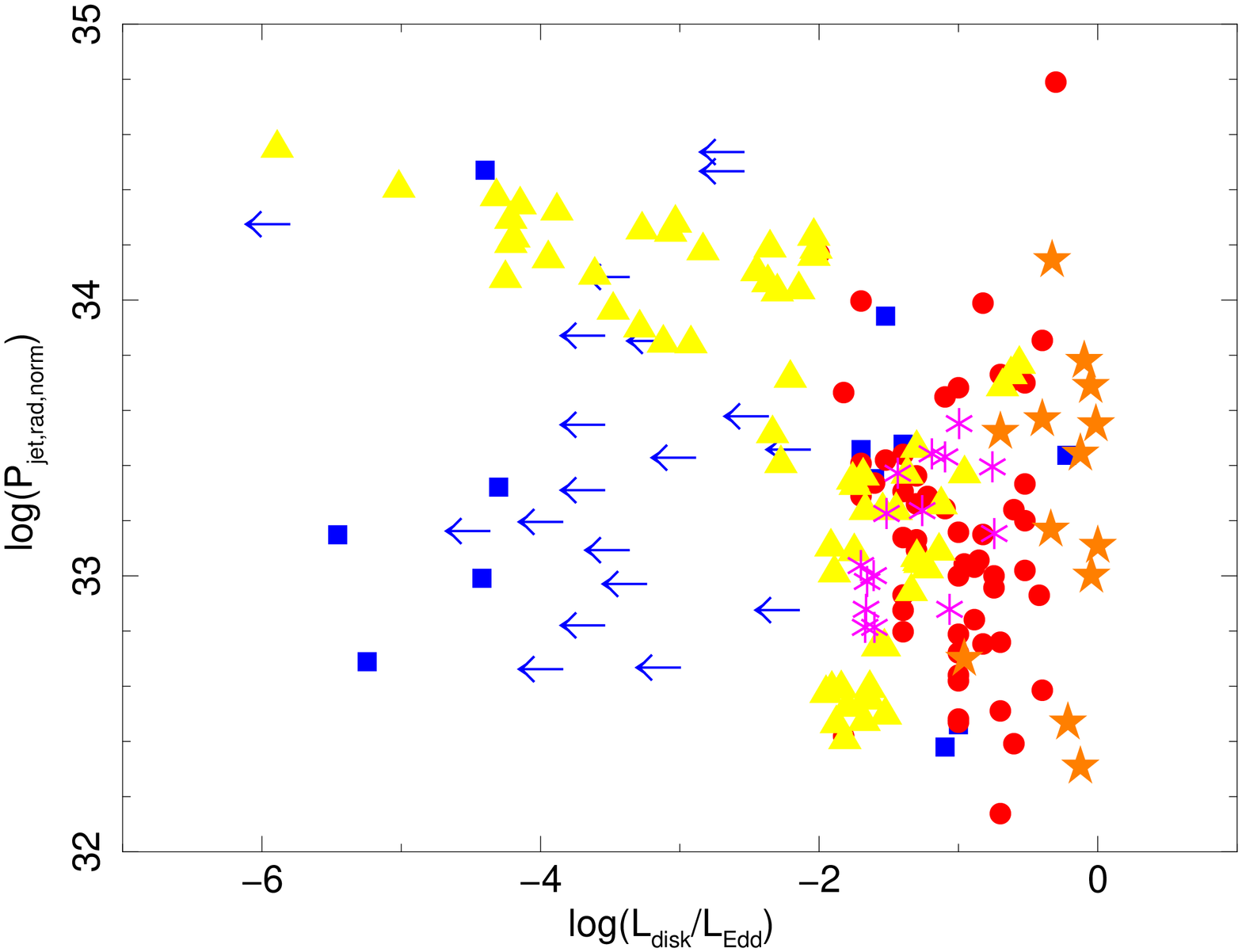}
\caption{({\it left panel}) Radiative Jet Power vs Disk Luminosity. ({\it right panel}) Normalized Radiative Jet Power vs Normalized Disk Luminosity. Different symbols indicate different sources: stellar-mass black holes (yellow triangles), neutron stars (cyan asterisks), FSRQs (red circles), BL Lac Objects (blue squares or arrows, as upper limits), RLNLS1s (orange stars). Jet power has been normalized according to Eqs.~(\ref{eq3},\ref{eq4}). \label{f1}}
\end{figure}

In the case of Galactic binaries, I started from X-ray fluxes, although in different energy bands ($3-9$~keV, $2-10$~keV, ...), then I renormalized all the fluxes to the $2-10$~keV band. The X-ray flux is linked to the accretion disk emission, which contributes a little during the hard state and for almost all the emission in the high state (see the well-known Fig.~1 in Ref. \refcite{ZDZ}). To extract the disk luminosity I would need to multiply by a factor that is small in low state and large in high state. Therefore, I can simply take the $2-10$~keV flux as reference of the disk and there will be a small error in the normalization, but not in the trend. The resulting graph displaying the radiative jet power vs the accretion disk luminosity is shown in Fig.~\ref{f1} ({\it left panel}). {\it The novelty of RLNLS1s is evident: it gives to the AGN region the missing branch equivalent to that of neutron stars for Galactic binaries, i.e. the low mass branch\cite{FOSCHINI1,FOSCHINI2,FOSCHINI3}.} This is very important in the scaling of relativistic jet: indeed, without the discovery of powerful relativistic jets from low-mass AGN (i.e. RLNLS1s), it would not be possible to unify AGN and Galactic binaries. If the former required a mass threshold, while the latter did not, this implied that jets in AGN were different from those in microquasars. 
 
Now that the two populations (AGN and Galactic binaries) share similar distributions in the disk-jet luminosity plane, it is possible to unify them by renormalizing the different powers. The accretion disk luminosity can be normalized by using the Eddington limit, $L_{\rm Edd}=1.3\times 10^{38}(M/M_{\odot})$~erg~s$^{-1}$. In the case of the jet power, Heinz and Sunyaev\cite{HEINZ} suggested that the scaling depends on whether the accretion disk is radiation-pressure domminated or gas-pressure dominated (see also Refs. \refcite{MODERSKI,GHOSH} and Fig.~3 in \refcite{FOSCHINI1}). In the former case, the jet power can be scaled according to the mass of the central compact object only:

\begin{equation}
\log P_{\rm jet,rad} \propto \frac{17}{12} \log M
\label{eq3}
\end{equation}

In the latter case, it is necessary to take into account also the accretion rate. It is not easy to measure the accretion rate, because it is necessary to know the accretion efficiency, which in turn depends on the spin of the central compact object in the case of a standard accretion disk. Given the present lack of reliability in both efficiency and spin measurements, I decided to keep as reference the disk luminosity. A satisfactory agreement can be found with the following scaling formula:

\begin{equation}
\log P_{\rm jet,rad} \propto \frac{17}{12} \log M + \frac{1}{2}\log \frac{L_{\rm disk}}{L_{\rm Edd}}
\label{eq4}
\end{equation}

where the scaling factor $(1/2)$ of the normalized accretion disk luminosity has been derived from the present data. It is different from the factor indicated by Heinz and Sunyaev\cite{HEINZ}, which instead referred to the accretion rate. 
The final result obtained by scaling according to the Eqs.~(\ref{eq3},\ref{eq4}) is displayed in Fig.~\ref{f1} ({\it right panel}). I note that the Galactic black holes tends to have greater normalized jet powers as the normalized disk luminosity decreases, meaning that the $2-10$~keV luminosity tends to understimate the disk luminosity in the case of low accretion. The error is within the spread of all the sources, due to many other reasons, and, therefore, I decided to keep this conversion. 

To further reduce the vertical spread, it would be necessary to understand how the spin of the central compact object affects the measurements. It is known that the rotation can have impact on both the jet generation\cite{BZ,FOSCHINI6} and the disk efficiency, but the lack of reliable measures -- particularly for AGN -- makes it difficult to set reliable pivotal points. In the case of microquasars, there is an open debate on whether the spin matters in the jet generation\cite{FENDER2,NARAYAN}. However, I note that if the Blandford-Znajek\cite{BZ} theory is valid, then what is important is the difference between the angular speeds of the central compact object and of the magnetic field threading the event horizon\cite{FOSCHINI6} -- and not the black hole spin itself. These details will be understood as soon as reliable measurements of the angular speeds will be available.

Anyway, already at the present stage, it is possible to conclude that the observed jet and accretion disk powers of AGN and Galactic binaries can be scaled according to the Heinz \& Sunyaev's laws\cite{HEINZ}. {\it Conducive to this result was the discovery of powerful relativistic jets from RLNLS1s, which made it evident the existence of a secondary branch in AGN similar to what was already known in Galactic binaries.}

\section*{Acknowledgments}
I am grateful to G. Tagliaferri (INAF OA Brera), F. Aharonian, F. Rieger, and the Organizing Committee for kindly supporting my participation at HEPRO IV.


\end{document}